\documentclass[sigconf]{acmart}

\AtBeginDocument{%
  }

\usepackage{enumitem}

\newcommand{\name}[2][ProForm]{\textit{#1#2}}

\newcommand{\B}{\textcolor{black}}

\setcopyright{acmlicensed}

\copyrightyear{2025}
\acmYear{2025}
\acmConference[UIST '25]{The 38th Annual ACM Symposium on User Interface Software and Technology}{September 28-October 1, 2025}{Busan, Republic of Korea}
\acmBooktitle{The 38th Annual ACM Symposium on User Interface Software and Technology (UIST '25), September 28-October 1, 2025, Busan, Republic of Korea}\acmDOI{10.1145/3746059.3747628}
\acmISBN{979-8-4007-2037-6/2025/09}

\begin{document}

\title[ProForm: Solder-Free Circuit Assembly Using Thermoforming]{ProForm: Solder-Free Circuit Assembly Using Thermoforming}

\author{Narjes Pourjafarian}
\email{n.pourjafarian@northeastern.edu}
\orcid{0000-0001-5298-6797}
\affiliation{%
  \institution{Northeastern University}
  \city{Boston}
  \country{USA}
}

\author{Zhenming Yang}
\email{yang.zhenm@northeastern.edu}
\orcid{0009-0005-5856-1809}
\affiliation{%
  \institution{Northeastern University}
  \city{Boston}
  \country{USA}
}

\author{Jeffrey Lipton}
\email{j.lipton@northeastern.edu}
\orcid{0000-0003-0843-0999}
\affiliation{%
  \institution{Northeastern University}
  \city{Boston}
  \country{USA}
}

\author{Benyamin Davaji}
\email{b.davaji@northeastern.edu}
\orcid{0000-0002-6375-4120}
\affiliation{%
  \institution{Northeastern University}
  \city{Boston}
  \country{USA}
}

\author{Gregory D. Abowd}
\email{g.abowd@northeastern.edu}
\orcid{0000-0002-3408-587X}
\affiliation{%
  \institution{Northeastern University}
  \city{Boston}
  \country{USA}
}

\renewcommand{\shortauthors}{Pourjafarian, et al.}

\begin{abstract}
  Electronic waste (e-waste) is a growing global challenge, with millions of functional components discarded due to the difficulty of repair and reuse. Traditional circuit assembly relies on soldering, which creates semi-permanent bonds that limit component recovery and contribute to unnecessary waste. 
We introduce \name{}, a thermoforming approach for solder-free circuit prototyping. By encapsulating electronic components with pressure-formed thermoplastics, \name{} enables secure, reversible mounting without the need for solder or custom mechanical housings. This approach supports a wide range of substrates, including flexible, paper-based, and non-planar circuits, facilitating easy reuse, \B{replacement}, and rapid prototyping. 
We demonstrate \name{’s} versatility to support prototyping practices. We show that \name{ed} circuits exhibit good electrical performance and mechanical stability.  While motivated by a need for sustainable electronics practices, \name{} has other significant advantages over traditional soldering.

\end{abstract}

%
%



\begin{teaserfigure}
  \includegraphics[width=\textwidth]{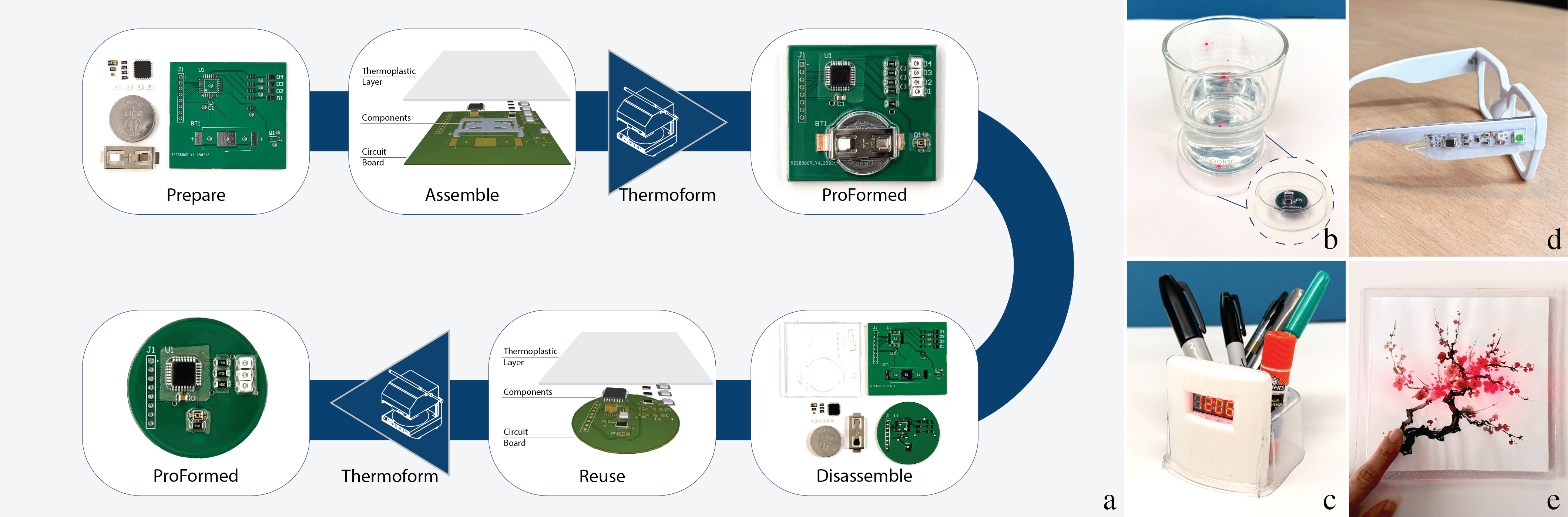}
  \caption{\name{}—A thermoforming approach for solder-free prototyping of electronic circuits. (a) The process involves assembling SMD components on a PCB and thermoforming a protective thermoplastic layer to secure them. The \name{ed} circuit is immediately functional. Later, the thermoplastic layer can be removed to enable the reuse of components. (b, c) \name{} facilitates seamless integration of electronics into interactive physical objects, (d) without compromising aesthetics. (e) \name{} supports alternative circuit substrates, such as paper.}
  \Description{}
  \label{fig:teaser}
\end{teaserfigure}

\maketitle

\section{Introduction}

Electronic waste (e-waste) has become a critical global challenge, with approximately 62 billion kg of e-waste generated worldwide in 2022 \cite{balde2024global}. This alarming figure results from inefficiencies in the repair, recovery, and reuse of electronic components. Current practices often require manual disassembly, desoldering, and separation of electronics to recover individual components. These processes are labor- and energy-intensive, as well as impractical at scale, discouraging efforts to extend the lifecycle of electronic devices. Consequently, when a single component fails, entire devices are discarded, including any remaining functional components, adding to the growing burden of landfills.

Addressing this issue through biodegradable, reusable, and repurposed electronics offers a significant opportunity to reduce e-waste and its environmental impact. In response, researchers have explored modular design \cite{Phidgets_UIST01,Bitblox_RoboResearch_2019,littleBits_TEI09,net-gadgeteer_Springer_2012}, biodegradable devices \cite{Degrade-to-Function_Lining_UIST24,biomaterials_UbiComp19, Mycelium-Artifacts_DIS19, Myco-accessories_ISWC19}, and upcycling approaches \cite{Unmaking_Song_CHI21}. Modular electronics enable component reuse and repair without discarding entire systems, while biodegradable and upcycled devices emphasize sustainability. Despite these advances, the predominant method for prototyping electronics continues to use semi-permanent bonding of SMD components with solder to the PCB. Desoldering often results in components being damaged during disassembly, further limiting reusability.

We introduce \name{}, a significant advance in the process of assembling electronic circuits without the use of solder. 
\name{} uses a mature technology, thermoforming, in a novel way to support to securely mount electronic components, eliminating the need for soldering or complex mechanical housings. This technique offers several key advantages. First, it facilitates component reuse and \B{replacement}, allowing easy removal and replacement without damaging the PCB or electronic devices or requiring desoldering. Second, it provides broad material compatibility, supporting a range of substrates, including traditional PCBs, flexible circuits, paper-based electronics, and non-planar surfaces. Third, \name{} enhances rapid prototyping, simplifying the circuit assembly process for a wider audience, including hobbyists, educators, and researchers. Finally, the approach aligns with sustainability and circular design principles by promoting the reuse of electronic components, reducing e-waste, and extending the lifespan of electronic systems.

We describe the process of \name{ing} and provide examples of its advantages over soldering and the wide variety of prototyping activities it supports. The main contributions of this paper are:
\begin{itemize}
    \item a novel fabrication approach for solder-free electronic prototyping using thermoforming, enabling easy reuse, \B{replacement}, and modification of circuits;
    \item a scalable and adaptable assembly method that supports planar and non-planar surfaces, accommodating diverse circuit substrates such as conductive inkjet and copper foil circuits, paper-based electronics, and flexible circuits; and
    \item demonstrations and evaluations showcasing the effectiveness of our approach for solder-free prototyping, highlighting its impact on sustainability and electronics prototyping.
\end{itemize}

\section{Related Work}
Our contribution builds on prior work in electronics and sustainable prototyping.

\subsection{Prototyping Electronic Circuits} 

    Prototyping electronic circuits is an essential step in developing and refining interactive systems, enabling rapid iteration and testing. Traditional approaches often rely on PCB fabrication and soldered connections, which, while effective, can be time-intensive, limit flexibility, and reduce component reusability. Recent innovations have introduced novel tools and techniques that simplify the process, enhance adaptability, and accommodate diverse design needs.
    For fabricating customized interfaces, techniques such as inkjet printing \cite{Kawahara_UbiComp13_Instant_Inkjet_Circuits, Khan_Soft_Inkjet, Perumal_UIST15_Printem, Print-a-sketch_CHI22, Pourjafarian_RoboSketch_CHI23}, screen printing \cite{olberding_printscreen_2014}, hydrographic printing \cite{groeger_objectskin_2018}, spray printing \cite{wessely_sprayable_2020}, and laser patterning \cite{Savage_Midas, Groeger_LASEC, Fibercuit_Yan_UIST22} have enabled the creation of circuits directly on unconventional surfaces. To facilitate faster, hands-on prototyping, researchers have explored using conductive styluses and paint \cite{Mellis_Microcontrollers_as_Material, Buechley_Paints_Paper, Koya_UbiComp15_ConductAR, Pourjafarian_BodyStylus_2021}, copper tape \cite{Savage_Midas}, or functional stickers \cite{cheng_silver_2020} that can be directly applied to real-world objects. 
    For augmenting electronic components onto these prototypes, existing approaches include applying conductive epoxy \cite{review-epoxy_Journal_2020}, which often makes it difficult to remove components without damage, and z-tape \cite{z-tape}, which requires external force to maintain reliable connections.

    Modular electronics provide a complementary approach to prototyping, emphasizing solder-free assembly and component reusability. Early platforms such as Phidgets \cite{Phidgets_UIST01} pioneered modular prototyping by offering sensor modules that could be easily integrated into designs. Subsequent innovations have expanded this concept to include plug-and-play platforms \cite{net-gadgeteer_Springer_2012, SoftMod_TEI20}, magnetic snapping modules \cite{littleBits_TEI09}, sticky modules \cite{Circuit-stickers_CHI14}, interconnecting blocks \cite{Bitblox_RoboResearch_2019}, and morphable sensor modules designed for 3D prototypes \cite{MorphSensor_UIST20}. More recent projects, such as BitBlox \cite{BitBlox_IDC16}, CurveBoards \cite{CurveBoards_CHI20}, and FlexBoard~\cite{FlexBoard_CHI23}, have further enhanced breadboard usability, enabling solder-free prototyping on even non-planar surfaces.
    These advancements enhanced the prototyping process. 

    Despite advancements in modular and solder-free prototyping, existing methods often require specialized components, compromise reliability, or limit compatibility with diverse circuit substrates. \name{} overcomes these challenges by introducing a thermoform-\\ing-based encapsulation approach, enabling secure, solder-free mounting on rigid, flexible, and paper-based circuits. Unlike adhesives or snap-in connectors, it ensures mechanical stability, easy component reuse, and reliable electrical performance, promoting sustainable and adaptable fabrication.

\subsection{Toward Sustainable Electronics Prototyping}

    The rapid growth of electronic waste (E-waste) has driven HCI researchers to explore sustainable alternatives for electronics prototyping, focusing on minimizing waste, extending device lifecycles, and enabling circular design practices. One promising direction involves the use of bio-based and biodegradable materials \cite{Degrade-to-Function_Lining_UIST24}, such as mycelium-based electronics~\cite{biomaterials_UbiComp19, Mycelium-Artifacts_DIS19, Myco-accessories_ISWC19}, bio-based~\cite{Bioplastics_UIST22, Functional-Destruction_CHI23, VIM_Song_CHI23, EcoThreads_Jingwen_CHI24} and recyclable polymers~\cite{Recy-ctronics_arXiv24, zhang2024recyclable, DissolvPCB_UIST25}, and cellulose- and paper-based interfaces~\cite{Decomposable_CHI22, Biohybrid-Devices_UIST23, SwellSense_Tingyu_CHI23}. These materials offer decomposable and recyclable solutions, reducing the environmental footprint of discarded electronics. Complementing this line of research are efforts in unmaking and upcycling \cite{Unmaking_Song_CHI21}, which focus on extending device lifespans by reusing and repairing components~\cite{ecoEDA_Pedro_UIST23, Unmaking_Pedro_ToCHI24, Breakdown_Jackson_CHI14, reuse-of-e-waste_Kim_CHI11, FabricatINK_Hanton_CHI22, Multifunctional-E-Textiles_Irmandy_IMWUT24} and dismantling smart textiles into reusable parts~\cite{Unfabricate_Devendorf_CHI20}.
    
    Meanwhile, efforts to recycle components from existing PCBs have focused on both DIY and industrial methods. Tools such as solder wicks, pumps, and hot air guns, along with industrial processes like thermal dismantling (e.g., hot air rework stations, IR reflow ovens) and chemical methods (e.g., solvent baths) \cite{oke2024discarded, Dismantling_Sustainability21}, provide standardized recovery techniques. However, these methods are often costly, time-intensive, and environmentally harmful, with components likely to sustain damage during recovery. 
    
    Recent work, the SolderlessPCB \cite{SolderlessPCB_Yan_CHI24}, explored an alternative to soldering by introducing 3D-printed housings that mechanically hold SMD components on the PCB. The housing was held together with screws that were easily tightened and loosened. While a promising approach for the reuse of components, this method has several limitations. The 3D-printed housing is unique to the PCB layout, making it impractical for complex circuits and requiring the design and fabrication of a specific new housing for each new circuit design.  It is also extremely difficult to design housing for a non-planar circuit, and offers no opportunity for a flexible circuit. 
    
    These challenges underscore the need for a simpler, sustainable solution that facilitates rapid, iterative prototyping for efficient component reuse, as well as supporting diverse surface geometries. This paper introduces \name{}, a thermoforming-based, solder-free approach that enables secure attachment, reuse, and replacement of components on both conventional and non-planar PCBs, aiming to advance sustainable electronics prototyping.

\section{WHAT IS PROFORMING}

\name{} is a thermoforming-based approach for solder-free prototyping of electronic circuits, enabling easy reuse and replacement of components. The term \name{ing} (\textbf{pro}totyping through thermo\textbf{form}ing) refers to the process of mounting electronic components onto a PCB, thermoforming a protective layer, and trimming excess thermoplastic. While it was designed to support a sustainability goal of reversible manufacturing and reuse, we will demonstrate how it has other advantages over traditional soldering. \name{} simplifies the assembly and disassembly of electronic components on both planar and non-planar surfaces, supporting various circuit fabrication methods, including conductive inkjet and screen printing, copper foil circuits, laser-etched substrates, paper-based electronics, and flexible circuits, in addition to conventional PCBs.

\subsection{Preparing the Circuit} \label{preparing}
   
    To prepare the circuit, standard PCB design tools can be used. It is recommended to include test points \B{in the PCB layout} to facilitate debugging and troubleshooting. In addition, integrating small vent holes (\(\sim 0.5~mm\)) into the design improves air circulation during thermoforming, enhancing material conformity and overall forming quality.
    
    To mount SMD components to the circuit \B{and to keep the components steady during handling and thermoforming}, we used \textbf{aniso-\\tropic conductive film} (z-tape).
    Z-tape securely holds components in place and compensates for minor height variations, ensuring consistent pin-to-pad contact.
    \B{Z-tape can be applied per component or as larger strips across sections of the PCB.}    
    \emph{For optimal contact resistance, z-tape must be compressed with uniform pressure, such as that applied by \name{}, along the bond line}~\cite{z-tape}. \B{Alternative attachment strategies are discussed in} Section~\ref{discussions}.

\subsection{Thermoforming}
    Thermoforming is a manufacturing process widely used to shape thermoplastic sheets into specific forms. It involves several sequential steps: 
    \begin{enumerate}
        \item[Heat] A thermoplastic sheet is heated to a temperature that makes it pliable.
        \item[Form] Once softened, the plastic is stretched over a mold to achieve the desired shape.
        \item[Cool] The material is then cooled to solidify and retain the mold's contours.
        \item[Trim] Any excess material is removed to produce a clean, finished part.
    \end{enumerate}
    Common techniques in the forming step include \textbf{vacuum forming}, where a vacuum pulls the softened material tightly against the mold, and \textbf{pressure forming}, which uses positive air pressure to achieve intricate details and sharper features.
    For \name{}, we used the pressure-forming technique,\footnote{\url{https://mayku.me/multiplier}} which is compatible with both commercial and DIY pressure-forming machines, and used PETG \cite{vivak-PETG} as the thermoplastic layer due to its strength, lightweight properties, transparency, and recyclability.
    
\begin{figure}
      \includegraphics[width=\columnwidth]{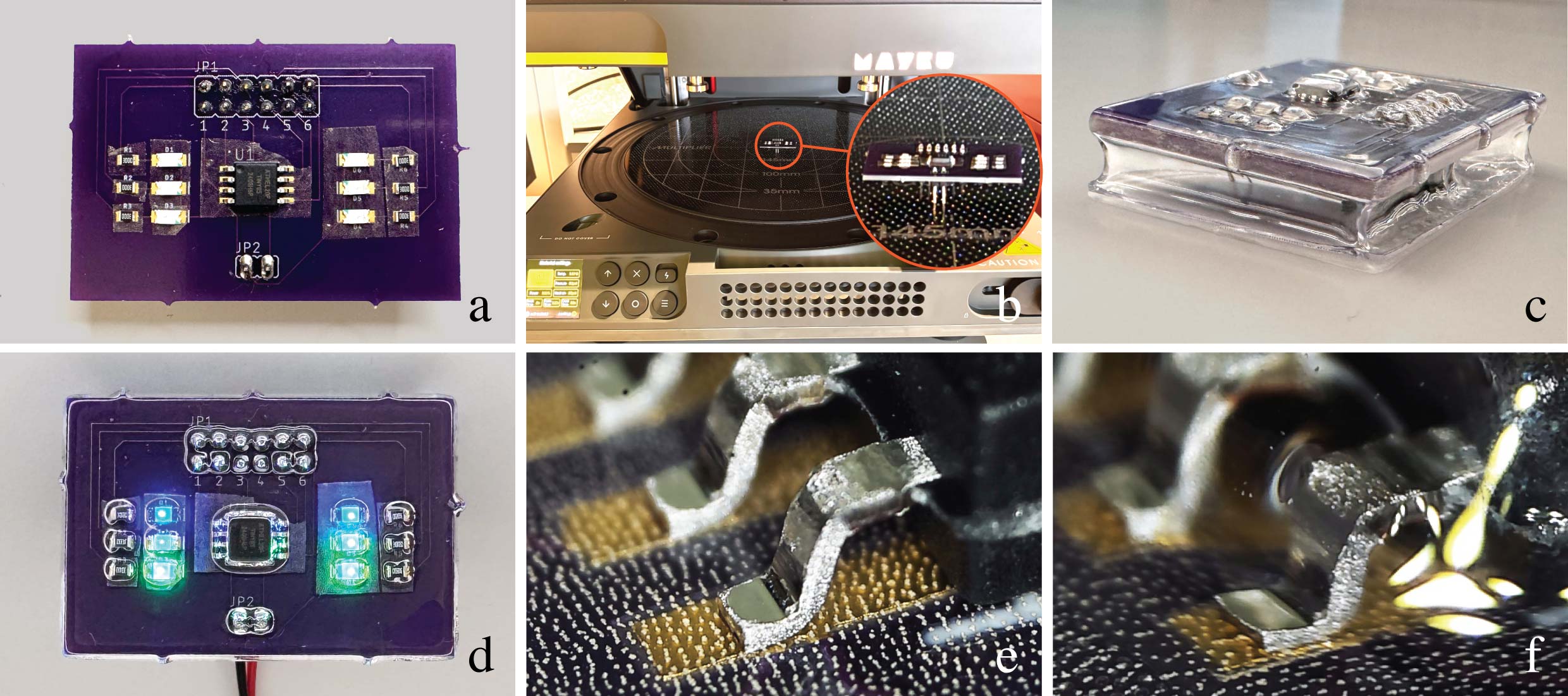}
      \caption{ProForming process: (a) Mounting SMD components on the PCB using Z-tape, (b,c) thermoforming while elevating the PCB to ensure proper material wrapping, and (d) trimming excess thermoplastic. Close-up views showing the connections between the PCB and an electronic component’s pads (e) before and (f) after thermoforming.}
      \Description{}
      \label{fig:thermoforming}
    \end{figure}

    After assembling the PCB and SMD components (Figure~\ref{fig:thermoforming}a), it is placed in the forming machine (Figure~\ref{fig:thermoforming}b).
    To ensure the thermoplastic sheet wraps securely around the board's edges, the PCB must be elevated from the machine's flat surface (Figure~\ref{fig:thermoforming}b, c).
    Elevation can be achieved using pin headers, small spacer objects, or temporary supports positioned under the PCB (Figure~\ref{fig:thermoforming}b-inset).    
    The thermoforming process lasts 210 seconds.
    Afterward, excess thermoplastic is trimmed using scissors or a hot knife (Figure~\ref{fig:thermoforming}d).
    Figures~\ref{fig:thermoforming}e and \ref{fig:thermoforming}f provide close-up views of the connections between the PCB pads and component pins (SOIC package), secured with Z-tape, before and after thermoforming.

\subsection{\B{Reusing and Replacing}} \label{repair-Reuse}
\begin{figure}
      \includegraphics[width=\columnwidth]{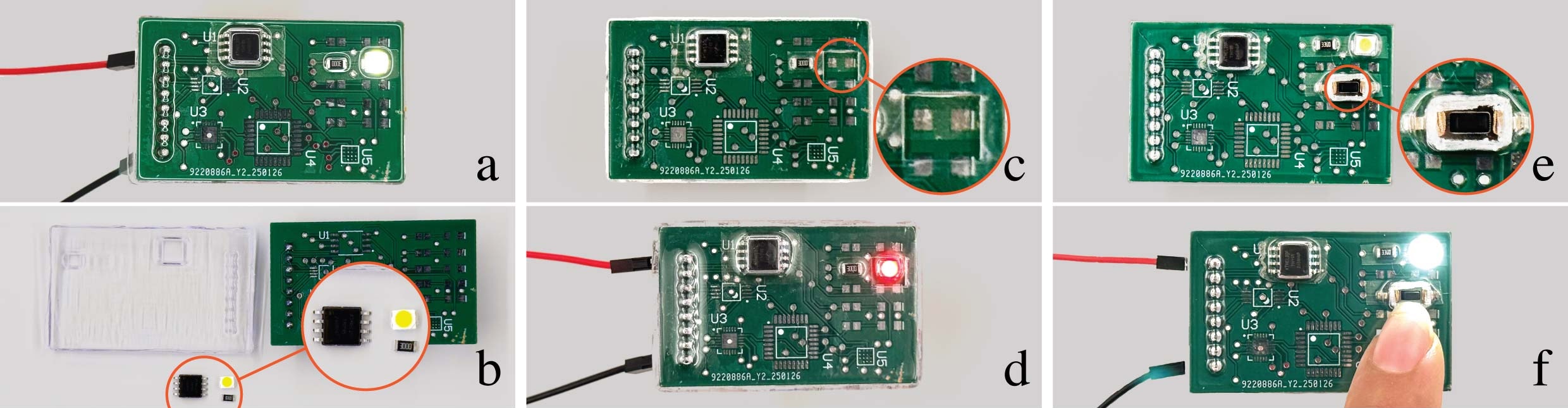}
      \caption{(a, b) Removing the thermoplastic layer to access and repurpose components. (c, d) Cutting around a component (e.g., LED) for replacement and re-thermoforming afterward. (e, f) Providing access to interactive components (e.g., push button) while preserving structural integrity and electrical connections.}
      \Description{}
      \label{fig:reusing}
    \end{figure}

    To \textbf{reuse} components after \name{ing} (Figure~\ref{fig:reusing}a), the thermoplastic layer can be removed using scissors, hot knife
    \cite{hot-knife}, or laser cutter. Once separated, the components are fully accessible and ready for reuse (Figure~\ref{fig:reusing}b).
    For \textbf{replacing} a specific component, the thermoformed layer around the component can be cut using a hot knife or laser cutter (Figure~\ref{fig:reusing}c). After the replacement, the board can be \name{ed} again (Figure~\ref{fig:reusing}d).
    Additionally, for providing \textbf{access} to interactive components such as push buttons or connectors, careful cutting with a laser cutter or hot knife is recommended in order to expose the desired element (Figure~\ref{fig:reusing}e). It is essential to ensure that the thermoplastic layer remains intact around other components to maintain reliable electrical connections (Figure~\ref{fig:reusing}f).

\subsection{Thermoforming Parameters}
    The effectiveness of \name{} is influenced by key thermoforming parameters, including material selection, sheet thickness, temperature, and pressure settings.

\vspace{2mm}    
\noindent \textbf{Materials and sheet thickness:} 
    In selecting a thermoformable material for \name{}, we prioritized recyclability, transparency, and semi-rigidity to support both visual inspection and mechanical stability. Polyethylene Terephthalate Glycol (PETG) was selected due to its low forming temperature, high impact resistance, ease of fabrication, and compatibility with standard desktop thermoforming tools \cite{Mayku}. PETG also offers fast processing times, is highly recyclable, and is cost-effective (1.08~USD per A4-sized sheet) \cite{vivak-PETG}.

    To identify the optimal sheet thickness, we \name{ed} fifteen zero-ohm resistors (size: 1206)—five per thickness—using 0.5~mm, 1~mm, and 1.5~mm PETG sheets. We then measured the electrical resistance across connections post-thermoforming. 
    The 0.5~mm sheet produced inconsistent results (average resistance: 15.8~$\Omega$, standard deviation: 7.39~$\Omega$). In contrast, both the 1~mm and 1.5~mm sheets maintained reliable performance (1~mm: Avg = 1.13~$\Omega$, SD = 0.18~$\Omega$; 1.5~mm: Avg = 0.85$\Omega$, SD = 0.44~$\Omega$). Given the comparable electrical performance, we selected 1~mm PETG as the default thickness to reduce material use and lower manufacturing and recycling costs.

\vspace{2mm}    
\noindent \textbf{Thermoforming settings:} 
    Following manufacturer recommendations \cite{vivak-PETG}, we set the forming temperature to 160°C. We tested pressures between 55 psi and 63 psi (the machine’s maximum limit) and observed that pressures above 58 psi produced sharper details and more reliable electrical connections. The forming cycle lasts 120 seconds, followed by a 90-second cooling phase to ensure proper solidification and structural integrity.

\subsection{Advantages Over Soldering}
Thermoforming offers several advantages over traditional soldering in circuit prototyping.

\begin{figure}
  \includegraphics[width=\columnwidth]{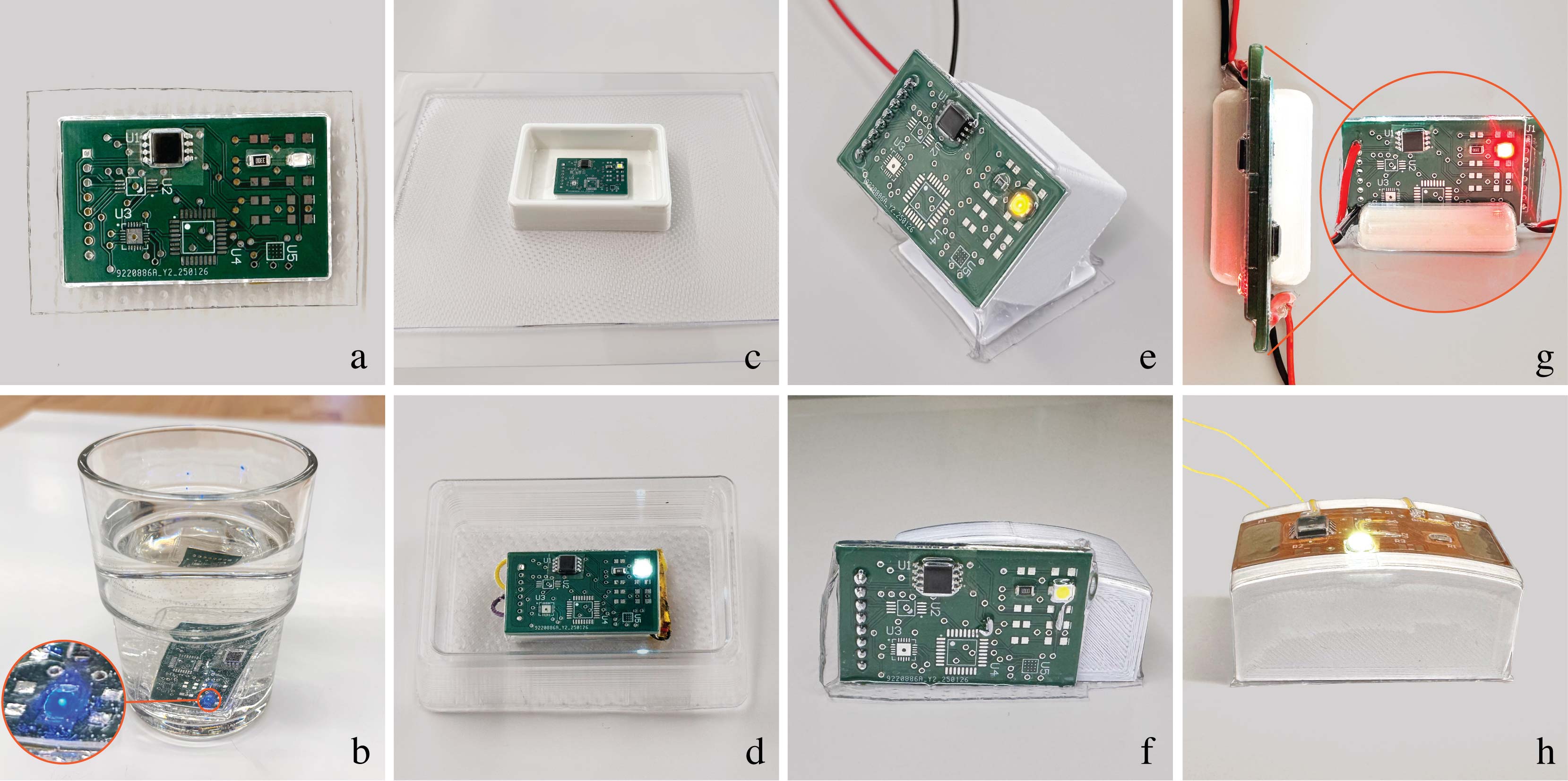}
  \caption{(a, b) Thermoforming both sides of a PCB achieves full encapsulation. (c, d) Integrating PCBs into molds allows them to become part of 3D physical objects. (e, f) Tilted and vertically positioned boards, (g) boards with components on both sides (top and side views), and (h) curved boards are supported.}
  \Description{}
  \label{fig:sealing}
\end{figure}

\vspace{2mm}
\noindent \textbf{Encapsulation and sealing:}
    \name{} enables circuit encapsulation, protecting boards from corrosion, dust, and moisture, while supporting waterproof applications. 
    To fully seal a board, the process begins by placing the board flat (without elevation) in the forming machine, thermoforming one side, and trimming excess thermoplastic, leaving a few millimeters for bonding with the second layer (Figure~\ref{fig:sealing}a). The board is then flipped, and the other side is thermoformed, creating a fully enclosed, sealed structure (Figure~\ref{fig:sealing}b).

\vspace{2mm}
\noindent \textbf{Integration:}
    \name{} allows circuits to be directly integrated into physical objects, eliminating the need for separate casings. This involves fabricating a mold (e.g., 3D-printed), placing the circuit board inside, and thermoforming them together (Figure~\ref{fig:sealing}c). Once the process is complete, the mold is removed, leaving the circuit securely embedded within the structure (Figure~\ref{fig:sealing}d).

\vspace{2mm}
\noindent \textbf{Angled, two-sided, and non-planar surfaces:}
    \name{} also supports thermoforming at different angles, such as tilted or vertical orientations (Figure~\ref{fig:sealing}e, f), and even allowing components on both sides of a PCB to be secured simultaneously (Figure~\ref{fig:sealing}g). Moreover, it accommodates curved surfaces, provided the curvature allows direct contact between component pins and the surface (Figure~\ref{fig:sealing}h).

\vspace{2mm}
\noindent \textbf{Adapting SMD components for prototyping:}
    A common approach to prototyping electronic circuits involves assembling the circuit on a breadboard with through-hole (THT) components. 
    However, this method relies on the availability of THT components, which is not always practical. Many electronic components exist only in surface-mount (SMD) format, and direct substitution between THT and SMD versions may be challenging due to differences in size, footprint, and electrical characteristics.
    Additionally, certain prototyping scenarios require the temporary attachment of SMD components without permanently bonding them to a PCB. For example, space constraints may prevent the inclusion of programming headers, necessitating external programming of an SMD microcontroller before integrating it into the final circuit. Moreover, verifying the functionality of previously used SMD components is not straightforward with conventional prototyping tools.
    
\begin{figure}
      \includegraphics[width=\columnwidth]{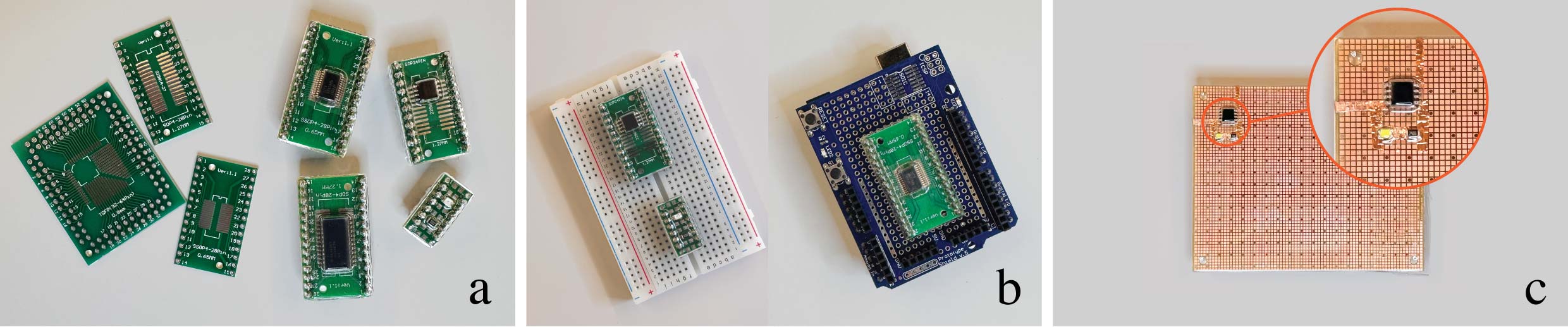}
      \caption{(a) \name{} enables SMD-to-THT conversion using adapter PCBs, (b) compatible with breadboards and protoboards. (c) Prototyping with SMD components directly on SMT pads is also supported.}
      \Description{}
      \label{fig:adapting-SMD}
    \end{figure}

    With \name{}, SMD components can be integrated into THT prototyping, programmed externally, and reused efficiently. A small adapter PCB facilitates SMD-to-THT conversion (Figure~\ref{fig:adapting-SMD}a), enabling component testing on breadboards and protoboards (Figure~\ref{fig:adapting-SMD}b).
    Additionally, \name{} supports rapid prototyping with SMD components directly on SMDpads \cite{SMT-pads}. The gaps between components can be bridged using solder or copper foil to establish connections (Figure~\ref{fig:adapting-SMD}c).
    Once validated, the thermoformed layer can be removed, allowing components to be reused for future designs.

\section{Performance}
This section presents an evaluation of \name{} through experimental validation and discussion of key findings. We assess its impact on circuit performance and mechanical robustness. The results provide insights into the feasibility, reliability, and scalability of \name{} for sustainable electronics prototyping.

\begin{figure*}
  \includegraphics[width=\textwidth]{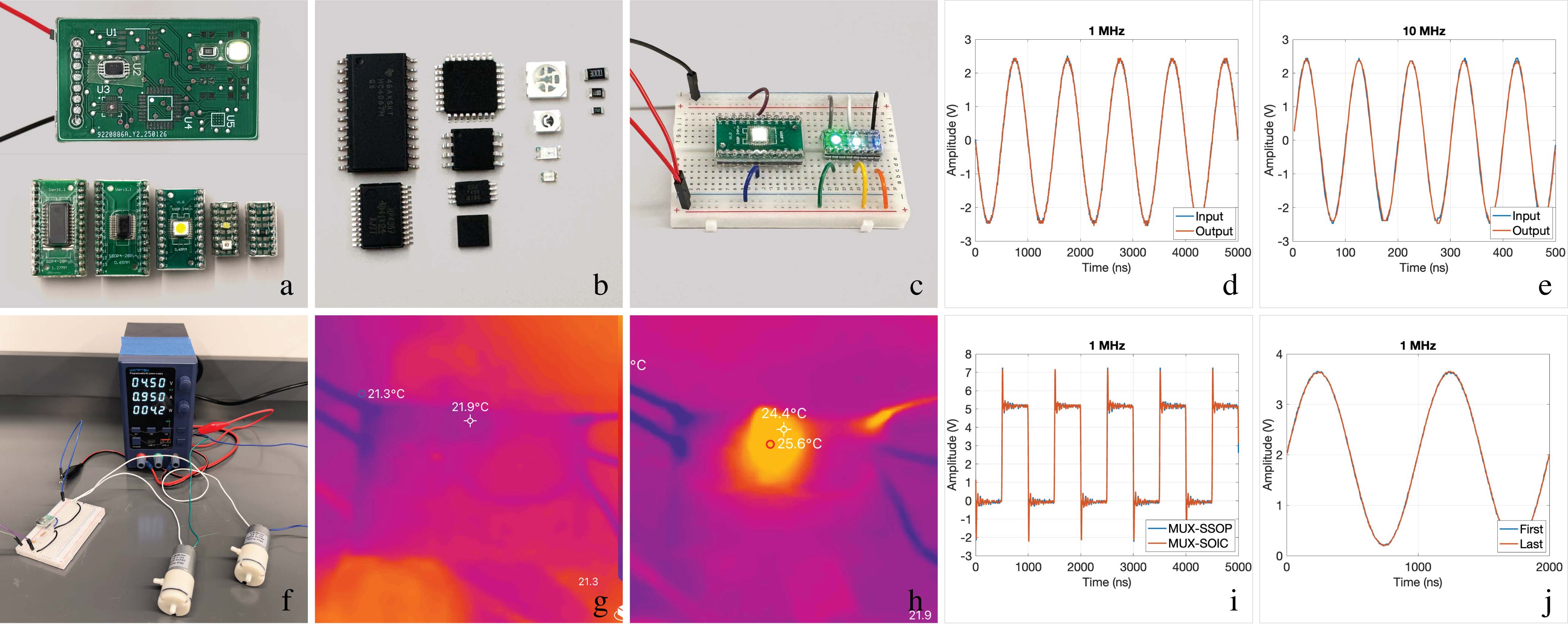}
  \caption{(a) Top: the PCB used for validating various ATtiny SMD packages. Bottom: adapter PCBs are used to test different IC packages, LED types, and resistor sizes \B{(Sections~\ref{resistance}, \ref{package}, \ref{reusability})}. (b) A range of SMD component packages and sizes is used for validation \B{(Sections~\ref{package})}. 
  (c) LEDs of various sizes are under test \B{(Sections~\ref{package})}.
  (d, e) Recorded input and output waveforms across a zero-ohm resistor \B{(Sections~\ref{frequency})}.
  (f) Experimental setup for high-current tolerance testing \B{(Sections~\ref{high-current})}. 
  (g, h) Thermal images of the high-current circuit before and after the experiment \B{(Sections~\ref{thermal})}.   
  (i) Comparison of PWM output signals from two different multiplexer packages \B{(Sections~\ref{package})}. 
  (j) Recorded output signals from a multiplexer after the first and tenth thermoforming cycles \B{(Sections~\ref{reusability})}.}
  \Description{}
  \label{fig:eval1}
\end{figure*}

\subsection{Circuit Performance} 

Does a \name{ed} circuit perform as well as a soldered circuit?
To evaluate the electrical reliability of \name{}, we conducted several experiments examining connection integrity, long-term stability, and reusability—key factors affecting circuit performance in prototyping.

\subsubsection{Connection integrity: electrical resistance} \label{resistance}

    We investigated how \name{} affects electrical connection quality by measuring the resistance of zero-ohm resistors across two common SMD sizes (0603 and 1206). For each size, five resistors were mounted on separate PCBs using z-tape and then thermoformed (Figure~\ref{fig:eval1}a, bottom right). Resistance measurements were taken via pin headers using a digital multimeter (Keithley DMM6500).
    \B{Across all samples, the average resistance was 1.32 $\Omega$ (1206: 1.13 $\Omega$, or 1.75$\Omega/\square$; 0603: 1.51 $\Omega$, or 2.01 $\Omega/\square$), with a standard deviation of 0.34 $\Omega$.}     
    For comparison, a 1205 resistor attached with z-tape but without thermoforming initially exhibited a resistance of 2.2k~$\Omega$ and, over time, failed entirely.

\subsubsection{Connection integrity: high-frequency signal} \label{frequency}
    To assess whe-\\ther \name{} affects signal integrity, we applied sinusoidal waves ranging from 100~KHz to 10~MHz to the zero-ohm resistors tested in the previous experiment. A function generator (BK Precision 4053) supplied input signals, while an oscilloscope (Keysight EDUX1002G) recorded the output waveforms.   
    Testing across both SMD resistor sizes confirmed no observable waveform attenuation in \name{ed} circuits. Figure~\ref{fig:eval1}d, e illustrates the input and output signals (sine wave, 5~Vp-p, 1~MHz and 10~MHz) applied to a 1206 zero-ohm resistor.

\subsubsection{Connection integrity: high-current tolerance} \label{high-current}
    To evaluate \name{'s} performance in high-current applications, we designed a circuit using a Power MOSFET (AOD452) to drive two DC motors (ZR370-02PM, 500~mA) (Figure~\ref{fig:eval1}f). The MOSFET was \name{ed} on an SMD adapter PCB.
    The motors were controlled via a PWM signal (50\% duty cycle, 0.5 Hz), toggling on and off every second over a one-hour period. The current flow was recorded every fifteen minutes, yielding an average current of 946~mA (STD: 21~mA).
    \B{To further assess continuous load performance, we repeated the test using two type-130 DC motors (1.5–6~V) operated continuously for one hour under identical recording conditions. The results again indicated consistent current flow with no observed degradation in connectivity or circuit performance (Avg: 984~mA, STD: 11~mA).}
    However, further evaluation is needed to characterize performance under higher currents and extended usage durations.

\subsubsection{Connection integrity: package type}\label{package}
    To examine how package type affects connection reliability, we tested five common SMD packages: SOIC (ATtiny85), TSSOP (ATtiny45), TQFP (ATtiny828), QFN (ATtiny85), and BGA (ATtiny84a). Each microcontroller was \name{ed} alongside a resistor and LED, then programmed via header pins to validate connections (Figure~\ref{fig:eval1}a-top, b).
    
    Results confirmed that SOIC, TSSOP, TQFP, and QFN packages established reliable electrical and mechanical contacts (as shown in the supplementary video). However, the BGA package failed to maintain a stable connection due to \B{its limited pad size, which did not provide sufficient conductive contact area for effective z-tape adhesion.} Attempts to \name{} this package without z-tape resulted in component displacement during handling and thermoforming. Future work should explore alternative attachment strategies, such as enhancing surface friction without compromising electrical conductivity (more details in Section~ \ref{discussions}).

    To evaluate how the component size and pin density influence reliability, we tested a multiplexer (CD74HC) in SOIC and SSOP packages (Figure~\ref{fig:eval1}a-bottom left, b), applying a PWM signal (1~MHz, 50\%~duty cycle) across all I/O pins. Both package variants maintained stable connections (Figure~\ref{fig:eval1}i).
    Various LED sizes (0805, 3014, 3528, and 5050) were tested to assess their compatibility with \name{}. All LEDs remained functional after thermoforming (Figure~\ref{fig:eval1}c).
    In addition, zero-ohm resistors in 0603, 0805, and 1206 sizes were tested in Section~\ref{resistance} and Application examples, with all sizes maintaining functional connections.

\subsubsection{Stability: aging effect}
    We developed \name{} to support rapid prototyping and reusability, but we are interested in understanding how stable it is over time. We therefore performed some tests to evaluate the long-term stability of the resulting \name{ed} circuits. A \name{ed} PCB, consisting of an ATtiny85 microcontroller and six LEDs (Figure~\ref{fig:thermoforming}a), was continuously powered from the day of assembly.
    After 152 days, the board remained fully functional, demonstrating that thermoformed connections maintain electrical integrity and mechanical stability over extended periods. 

 \subsubsection{Stability: thermal stress} \label{thermal}
    To assess thermal effects under high-current operation, we recorded the surface temperature of the MOSFET during the first high-current test (Section~\ref{high-current}) using a thermal camera (TOPDON TC002C). Measurements were taken at the start (Figure~\ref{fig:eval1}g) and end of the test (Figure~\ref{fig:eval1}h).
    After one hour of continuous operation, the MOSFET exhibited a minor temperature increase of 3.7°C, indicating no significant overheating. Future iterations could improve thermal dissipation by including heat sinks. Post-thermoforming, excess thermoplastic around the heat sink can be trimmed without compromising component stability.

\subsubsection{Component reusability} \label{reusability}
    To assess the reusability of components after multiple thermoforming cycles, we tested a multiplexer (CD74HC, SSOP package, Figure~\ref{fig:eval1}a) under ten (dis)assembly cycles.
    The MUX was first \name{ed}, then tested using sinusoidal signals (1~MHz, 3.5~Vp-p). Next, the thermoplastic layer was removed, and the MUX was re-thermoformed onto a new board.
    Output waveforms were recorded for each iteration using an oscilloscope.
    Across all ten cycles, no measurable difference in output was observed. Figure~\ref{fig:eval1}j illustrates the output signals from Pin 9 (I0) after the first and last thermoforming cycles.

\subsection{Mechanical Robustness}
\name{} provides a protective layer that shields components from moisture, dust, mechanical impact, and handling stress. 

\subsubsection{Sealing and water resistance}
    To evaluate sealing performa-\\nce, we thermoformed a PCB on both sides, encapsulating an ATtiny85 MCU, an LED, and a coin cell battery (Figure~\ref{fig:sealing}a, b). The circuit was submerged in water for 24 hours. The circuit remained fully functional throughout and after the experiment.

\subsubsection{Drop Test}
    To assess mechanical robustness, we performed a drop test. The \name{ed} PCB in Figure~\ref{fig:eval1}a-top was dropped from heights ranging from 50 cm to 200 cm, at 50 cm intervals. We conducted three drops from each height; the PCB assembly remained fully functional throughout the entire series of drop tests.

\section{Applications and Use Cases}\label{applications}
To demonstrate the feasibility and versatility of \name{}, we present seven application examples that showcase how our approach enables repairing, reusing, and encapsulating circuits across diverse substrates—including paper, flexible, and rigid PCBs—in various orientations and curvatures. These examples highlight sustainability, rapid prototyping, and integration with everyday objects, illustrating \name{'s} potential for wearables, interactive devices, and DIY electronics.

    \begin{figure*}
          \includegraphics[width=\textwidth]{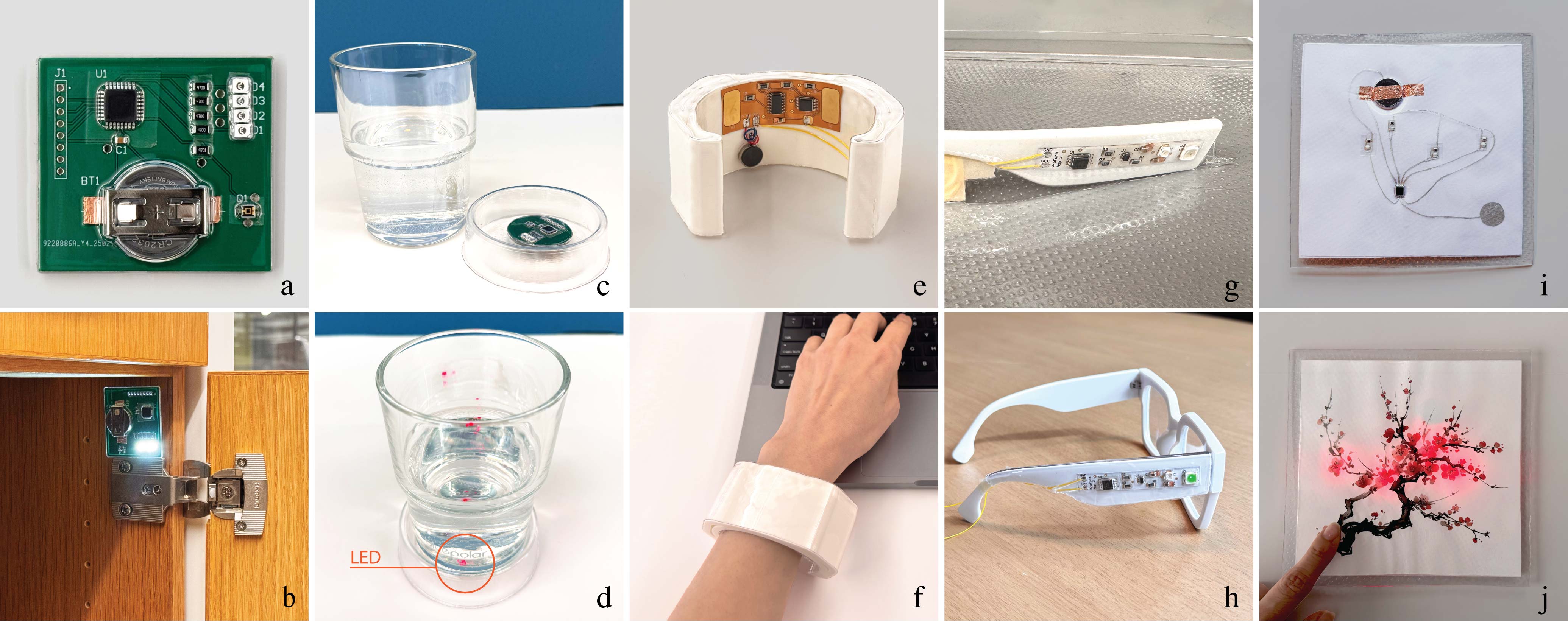}
          \caption{Reusing electronic components \B{(Section~\ref{app1})}: (a, b) a smart switch initially built for a cupboard; (c) repurposing the used components into a hydration reminder device, embedded in a cup holder base; (d) if the cup remains untouched, the LED turns red. Interactive wristband \B{(Section~\ref{app2})}: (e, f) a flexible circuit with a capacitive touch sensor and an actuator; the circuit is \name{ed} onto a concave surface of a 3D-printed wristband. 
          UV exposure detector \B{(Section~\ref{app3})}: (g, h) smart glasses with an integrated UV sensor and LED feedback; extensive exposure to UV light triggers a color change from green to red.  
          Interactive greeting card \B{(Section~\ref{greeting-card})}: (i) the hand-sketched circuit on one side of the paper and (j) the aesthetic layer on the other side; touching the capacitive sensor activates the LEDs.  
          }
          \Description{}
          \label{fig:app1}
    \end{figure*}

\subsection{Reusing Components: From Smart Switch to Hydration Tracker}\label{app1}

    \name{} supports component reuse, enabling easy modification and repurposing of circuit parts across different projects. We first assembled a smart cupboard switch designed to automatically turn on a light when the door opens. The circuit includes a microcontroller (ATtiny828), ambient light sensor (TEMT6000), four LEDs (1206 package), and a battery holder mounted on a 50 × 45~mm square PCB (Figure~\ref{fig:app1}a, b). The battery holder is attached using double-sided copper tape.

    After the light sensor malfunctioned, it was replaced by locally trimming the thermoplastic with a hot knife and re-\name{ing} the new sensor without removing the previous thermoplastic layer. 
    Later, the same components were repurposed into a hydration tracker. The circuit was transferred to a circular PCB (40~mm diameter) integrated into a cup holder. The LED colors were reconfigured and the device was used to indicate elapsed time since the last drink: when a cup is placed on the holder, a timer activates—a blue LED lights up initially, turning yellow after 30 minutes and red after an hour (Figure~\ref{fig:app1}c, d). Removing the cup resets the cycle.

\subsection{Circuit Integration on a Curved Surface}\label{app2}

    To explore \name{'s} suitability for curved surfaces, we developed an interactive wristband designed to help prevent prolonged typing strain (Figure~\ref{fig:app1}f). The user wears the wristband whenever they start working with a computer. The wristband detects continuous contact with the skin using the integrated capacitive sensors and provides haptic feedback as a rest reminder. The circuit, assembled on a flexible PCB, includes a capacitive touch controller (AT42QT1070), a mini vibration actuator, and an MCU (ATtiny85).
    The circuit is integrated into a 3D-printed wristband (the longest and shortest diameters: 65~mm and 50~mm). After \name{ing}, the thermoplastic covering the actuator is removed to allow unobstructed vibration (Figure~\ref{fig:app1}e).

\subsection{Smart Glasses with UV Exposure Detection}\label{app3}

    Wearable electronics must withstand environmental conditions such as dust, humidity, and impact while maintaining aesthetic integrity. \name{} enhances device resilience without compromising design.
    To demonstrate this, we developed UV exposure-detecting smart glasses with an embedded sensor in the temple. The circuit features a UV sensor (GUVA-S12SD), an RGB LED (WS2812B), and ATtiny85 (Figure~\ref{fig:app1}g).
    When the user is exposed to intensive UV light, the integrated LED transitions from green to red, signaling excessive exposure (Figure~\ref{fig:app1}h). The thermoplastic layer around the sensor can be removed for optimal accuracy.

\subsection{Durable Interactive Greeting Card} \label{greeting-card}

    Rapid DIY electronics prototyping—such as inkjet printing \cite{Kawahara_UbiComp13_Instant_Inkjet_Circuits}, hand-sketched circuits \cite{Mellis_Microcontrollers_as_Material}, and copper tape circuits \cite{Savage_Midas}—has enabled interactive designs on flexible substrates. However, attaching electronic components remains a challenge as soldering is impractical, and adhesives are unreliable.
    \name{} secures components onto these substrates, extending the lifespan of the electronics. 
    
    We created an interactive greeting card using a hand-sketched conductive circuit on standard paper (Figure~\ref{fig:app1}i). The card features an MCU (ATtiny85) and four LEDs. 
    When the sketched capacitive touch sensor is activated, the LEDs illuminate, brightening the blossoms of a printed tree (Figure~\ref{fig:app1}j). By securely encapsulating the components, \name{} extends the card's durability without compromising its interactive experience. However, it comes at the expense of the overall substrate flexibility.

\subsection{Pen Holder with a Clock} \label{app5}

\begin{figure}
          \includegraphics[width=\columnwidth]{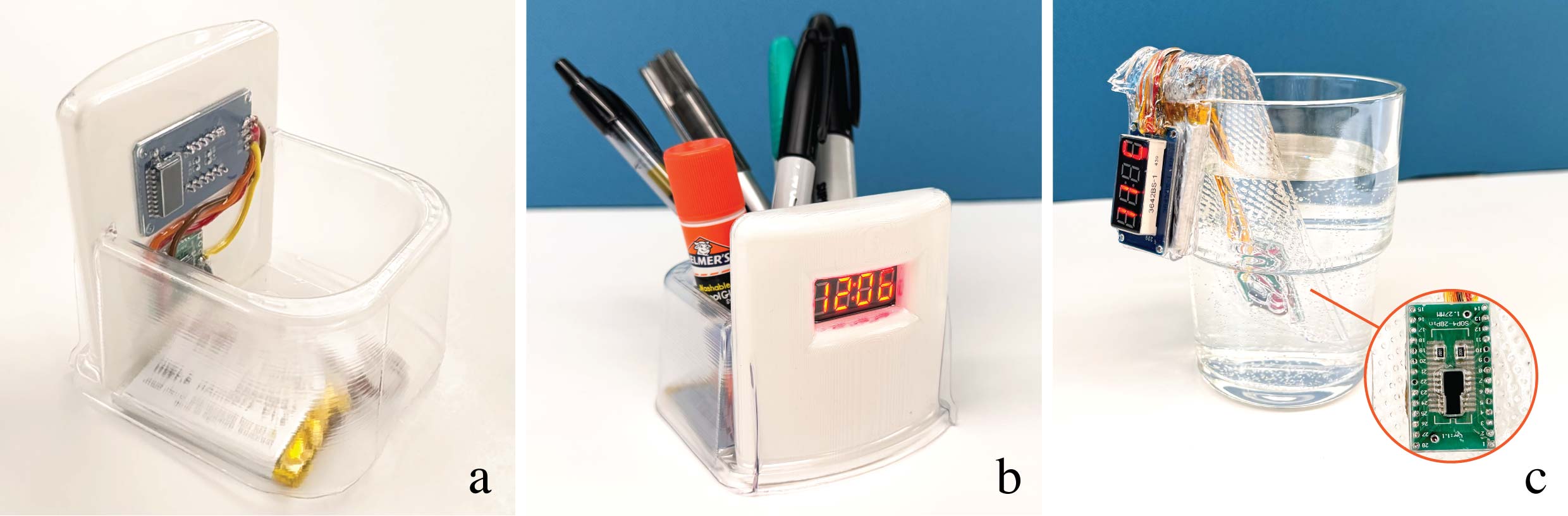}
          \caption{(a, b) A pen holder with an integrated clock: electronics are embedded within a 3D-printed mold, allowing functional affordances such as a designated battery compartment and pen storage. (c) A waterproof thermometer, sealed with thermoforming, designed for submersion in liquids.}
          \Description{}
          \label{fig:app2}
    \end{figure}

    To illustrate \name{'s} ability to support varied orientations, we developed a pen holder with an embedded digital clock (Figure~\ref{fig:app2}b). This example demonstrates how \name{} can secure existing electronic modules (e.g., seven-segment displays) without screws or adhesives, providing a seamless integration of electronics with a physical form factor.
    The circuit consists of a seven-segment display module and an ATtiny85 MCU.
    Instead of fabricating a new PCB, we used an adapter board, connecting the display to the MCU through wires. The MCU is programmed externally (using the \name{ing} approach), before being thermoformed into place. The electronic components are embedded within a 3D-printed holder (Figure~\ref{fig:app2}a), while the mold for the pen holder part is removed post-thermoforming, providing space for a battery.

\subsection{Waterproof Thermometer} \label{app6}
    To showcase \name{'s} sealing capabilities, we created a waterproof thermometer designed for submersion (Figure~\ref{fig:app2}c). The circuit includes a temperature sensor (LM75BD) and an ATtiny85 MCU, assembled on an SMD adapter PCB and connected to a seven-segment display module via wires, powered by a LiPo battery (3.7V, 1100mAh).
    All components were placed flat in the thermoforming machine and sealed on both sides before trimming excess thermoplastic. Finally, the circuit was bent using a heat gun to conform to the edge of a water glass.

\section{Technical Considerations and Future Directions}\label{discussions}
While \name{} provides a novel approach to solder-free circuit prototyping, several technical aspects can be optimized for improved reliability, scalability, and broader applicability. This section outlines key considerations and future directions.

\vspace{2mm}
\noindent \textbf{Surface preparation and component alignment: } 
    Reliable electrical connections depend on clean, flat PCB pads and uniformly aligned component leads. Surface contaminants, oxidation, or irregular pad topographies can introduce variability in contact resistance and compromise circuit performance. Proper surface preparation prior to component placement is therefore critical for maintaining connection integrity.

    Currently, z-tape is the primary method used to align and secure SMD components during thermoforming (Section~\ref{preparing}). \B{However, as it depends on sufficient contact area between the component pins and PCB pads to maintain conductivity, it is not compatible with packages that feature very small pad geometries} (Section~\ref{package}).

    \B{To explore alternative fixation strategies, we experimented with a DIY method using glycerin, applied between the PCB and the component body to introduce light friction and temporarily stabilize components. While effective for larger packages, this approach proved unsuitable for small-package components due to insufficient surface contact. We also tested carbon and silver-based conductive greases, applied via an SMD stencil. These materials offered electrical conductivity and moderate mechanical stability without forming a permanent bond. However, precise application is critical, as the grease can spread and potentially cause short circuits—particularly in high-density layouts.}

    Future work should investigate alternative fixation strategies that balance stability, conductivity, and reversibility. Promising directions include heat-releasable adhesives or customized solder mask patterns with embedded grooves or micro-indentations to guide and secure components during thermoforming, without compromising reusability or electrical performance.

\vspace{2mm}
\noindent \textbf{\B{Circuit debugging and repair: }}
    \B{\name{} shifts circuit repair from traditional wire-level modification to targeted, component-level intervention. After thermoforming, direct access to PCB traces is limited by the PETG layer; however, individual components can still be removed or replaced by locally trimming the thermoplastic using a hot knife or laser cutter (Section~\ref{repair-Reuse}). This selective disassembly approach preserves the surrounding connections, enabling partial repairs without requiring full circuit disassembly.}

    \B{To facilitate diagnostics and rework post-thermoforming, we recommend integrating test points into the PCB layout. These can be accessed via pogo pins, header connections, or, in the case of single-sided boards, from the exposed rear surface. Additionally, PETG’s optical transparency enables visual inspection of underlying components and pad alignment, further facilitating circuit debugging.}

\vspace{2mm}
\noindent \textbf{Pressure and contact optimization: }
    The current approach utilizes a maximum tested pressure of 63 psi, which effectively flattens slightly uneven pins but is insufficient for significantly deformed or misaligned ones. Higher-pressure settings should be explored to improve contact quality, particularly for components with irregular pin structures.

\vspace{2mm}
\noindent \textbf{Flexible circuits: }
    As we demonstrated with the example of the interactive greeting card (Section \ref{greeting-card}), \name{} supports flexible and paper-based circuits, but the thermoplastic we used, PETG, introduces rigidity, limiting the flexibility of the resulting prototype. Future research should investigate alternative thermoformable materials that maintain flexibility while ensuring reliable electrical connections.

\vspace{2mm}
\noindent \textbf{Through-hole components: }
    The current method primarily supports SMD components and does not yet accommodate through-hole technology (THT) components. As a result, connector headers and wires in the prototypes are soldered directly onto the PCB. Future work could explore alternative attachment methods, such as conductive pastes or flexible adhesives that remain pliable after curing. These materials could provide secure yet detachable THT connections, significantly broadening the range of prototyping possibilities with \name{}.

\vspace{2mm}
\noindent \textbf{Scalability for larger PCBs: }
    Scaling \name{} to larger circuit boards requires ensuring uniform pressure distribution across the surface. Future work should explore pressure mapping techniques to identify inconsistencies and improve reliability.
    Additionally, integrating structured venting or localized cutouts could enhance material conformity, preventing uneven surface contact and ensuring consistent electrical connections in large-scale designs.

\vspace{2mm}
\noindent \textbf{Component placement: }
    While \name{} is compatible with existing pick-and-place techniques, it also introduces an efficient workflow for assembling multiple PCBs. After thermoforming the first PCB, the thermoplastic layer can be removed and repurposed as a stencil for aligning components on subsequent boards. By placing a PCB with a z-tape layer onto this stencil, all components can be transferred simultaneously, ensuring accurate placement. This process could be repeated for batch fabrication, improving scalability and reducing alignment errors.

\vspace{2mm}
\noindent \textbf{Materials and environmental impact: }
    While PETG meets the recyclability requirements of \name{}, future work should prioritize the exploration of biodegradable and compostable alternatives to further reduce environmental impact. \B{As a widely available thermoplastic, PETG is compatible with both industrial recycling systems and decentralized tools such as desktop filament extruders, enabling localized material recovery and reuse. However, its petroleum-based origin limits its alignment with fully circular material lifecycles.} Advancing \name{} toward more sustainable fabrication will require identifying thermoformable substrates specifically engineered for electronic applications—materials that balance optical clarity, impact resistance, and conformability with biodegradability. Developing materials with tailored mechanical and electrical properties will ensure broader adoption of \name{} for sustainable electronics practices.

\vspace{2mm}
\noindent \textbf{Sustainability impact: }
    \B{\name{} was developed with sustainability as a core principle, targeting two key challenges in electronics fabrication: component disposability and the difficulty of circuit disassembly. By eliminating solder, \name{} enables the recovery and reuse of electronic components, thereby reducing material consumption and electronic waste.}

    \B{Quantifying the environmental impact of soldering is challenging, as it is deeply embedded within broader electronics assembly workflows. Rather than isolate soldering as a standalone process, we evaluate the implications of its removal—specifically, enabling component reuse through the addition of a thermoformed PETG layer. This trade-off is favorable: the embodied environmental impact of semiconductor components significantly exceeds that of the PETG used for encapsulation \cite{WANG202347}.}

    \B{While soldering processes—particularly reflow—consume energy and generate emissions \cite{geibig2005solders}, thermoforming contributes minimally to energy use, relying on moderate-temperature heating rather than high-intensity operations. Even modest reuse of integrated circuits or microcontrollers can offset the environmental cost of the PETG layer. Additionally, PETG waste generated during ProForming can be reclaimed using desktop recycling systems, significantly reducing lifecycle emissions compared to virgin material \cite{petg-Prusa}.}

\vspace{2mm}
\noindent \textbf{Broader impact: }
    Beyond facilitating component reuse, \name{} offers broader sustainability benefits by reducing packaging waste, shipping, and labor costs associated with manufacturing new components. By minimizing the need for permanent adhesives and solder, this approach can contribute to more sustainable electronics production and prototyping practices.

\vspace{2mm}
\noindent \textbf{Beyond prototyping: }
    \B{Although initially developed for circuit prototyping and component reuse, \name{}’s material versatility, mechanical robustness, and compatibility with batch workflows indicate strong potential for broader adoption in scalable manufacturing contexts. Realizing this potential requires key advancements: refining tooling to ensure uniform pressure distribution across larger boards and integrating component placement with automated pick-and-place assembly pipelines. Furthermore, investigating biodegradable thermoformable materials would support environmentally sustainable production. With these adaptations, \name{} could transition from a rapid prototyping method to a viable platform for circular, repairable, and low-impact electronics manufacturing.}

\section{Conclusion}
\name{} introduces a novel thermoforming-based approach for solder-free circuit prototyping, enabling secure, reversible mounting of electronic components without the need for solder or custom mechanical housings. By leveraging pressure-formed thermoplastics, our method supports a wide range of circuit substrates, including flexible, paper-based, and non-planar surfaces. This approach facilitates rapid prototyping while promoting sustainability by enabling easy reuse, \B{replacement}, and modification of circuits.
Through demonstrations and evaluations, we have shown that \name{} maintains good electrical performance and mechanical stability, making it a viable alternative to traditional soldering. Compared to existing modular and mechanical fastening techniques, \name{} offers a simpler, more adaptable solution that reduces material waste and extends the lifecycle of electronic components.
While primarily motivated by sustainability, \name{} also lowers the barrier to entry for prototyping and experimentation, making circuit assembly more accessible to a broader audience. 
By rethinking conventional circuit prototyping, \name{} contributes to the broader vision of sustainable, repairable, and modular electronics.

\begin{acks} 
We thank Canek Fuentes, Ruobing Bai, and Shaghayegh Mesforoush for their valuable insights throughout the development of this project.
\end{acks}

\bibliographystyle{ACM-Reference-Format}
\bibliography{sample-base}

\end{document}